\documentclass[12pt]{article}%
\usepackage{citesort}
\usepackage{amsmath}
\usepackage{graphicx}%
\usepackage{amsfonts}%
\usepackage{amssymb}
\usepackage{epsfig}
\newcommand{\bl}{\pmb{l}}
\newcommand{\bk}{\pmb{k}}
\newcommand{\bp}{\pmb{p}}

\newcommand{\be}{\begin{equation}}
\newcommand{\ee}{\end{equation}}

\newcommand{\bea}{\begin{eqnarray}}
\newcommand{\eea}{\end{eqnarray}}
\newcommand{\cM}{{\cal M}}
\begin{document}

\mbox{} \hfill DESY 07--015 \\
\mbox{} \hfill 13th February 2007\\[10mm]

\begin{center}
\textbf{\LARGE Exclusive $J/\psi$ and $\Upsilon$ hadroproduction \\[3mm] 
and the QCD odderon }\\
\vspace{2\baselineskip}
{\large
A.~Bzdak$^{a}$, L.~Motyka$^{a,b}$,
L.~Szymanowski$^{c,d,e}$ and J.-R.~Cudell$^{c}$
}
\\
\vspace{2\baselineskip}
${}^a$\,
{\it M.\ Smoluchowski Institute of Physics, Jagellonian University, Krak\'{o}w, Poland} \\[0.5\baselineskip]
${}^b$\, 
{\it II Institute of Theoretical Physics, Hamburg University, Hamburg, Germany} \\[0.5\baselineskip]
${}^c$\,
{\it Universit\'e  de Li\`ege,  B4000  Li\`ege, Belgium} \\[0.5\baselineskip]
${}^d$\,
{\it CPHT, {\'E}cole Polytechnique, CNRS, 91128 Palaiseau, France} \\[0.5\baselineskip]
${}^e$\,
{\it Soltan Institute for Nuclear Studies, Warsaw, Poland} \\[0.5\baselineskip]

\vspace{2\baselineskip}
\textbf{Abstract}\\
\vspace{1\baselineskip}
\parbox{0.9\textwidth}
{We study   $pp$ and $p\bar p$ collisions which lead to the exclusive production of $J/\psi$ or $\Upsilon$ 
from the pomeron--odderon and the pomeron--photon fusion. We calculate  scattering amplitudes of these processes in the lowest order approximation and 
in the framework of  $k_\perp$--factorization.  
We present estimates of cross sections
for the kinematic conditions of the Tevatron and of the LHC.
}
\end{center}

\section{Introduction}

It follows from the optical theorem that total cross sections of hadronic processes 
are driven by colour singlet exchanges in the $t$--channel. Thus, 
 pomeron exchange, 
characterized by an even charge parity, gives the dominant contribution to the sum of 
the direct and the crossed amplitudes for a given hadronic process.
 The exchange with the odd charge parity, i.e.\ that of the odderon, 
dominates the difference between these two amplitudes.
The concept of the odderon in the description of hadronic processes was introduced 
a long time ago \cite{Lukaszuk}. 
Although it is
 a partner of the pomeron, which is well known 
from the study of diffractive processes, the odderon still remains a mystery. 
As it differs from the pomeron only by its charge parity, one would expect,
 from the point
of view of general principles based on the analyticity and the unitarity of the $S$--matrix, 
that its exchange should lead to effects of a comparable magnitude to 
those coming from  pomeron exchange. However, the odderon still escapes experimental verification.

In  perturbative QCD, the pomeron is modeled by two interacting gluons  
in a colour-singlet state, whereas the odderon is described by an analogous 
system formed by three gluons. It is thus quite natural to expect that in 
hard processes the effects of odderon exchange ---  being suppressed 
by an additional power of the strong coupling constant $\alpha_{s}$ ---
are  smaller than similar contributions due to  pomeron exchange. 
This was confirmed by QCD studies of the diffractive exclusive 
$\eta_c$~production mediated by odderon exchange~\cite{eta,eta2,eta3,BBCV},
which led to rather small cross sections.
It was surprising, however, that  a non-perturbative description within the 
stochastic vacuum model of the similar exclusive process of the 
$\pi^0$~production \cite{stoch} gave a prediction which was disproved by  
experiment \cite{hera_odd}. It was then argued that the suppression of the $\pi_0$ 
photoproduction may emerge as a result of the chiral symmetry constraints
on the photon--$\pi_0$ coupling~\cite{chiral} or of the odderon absorption
by its coupling to the pomeron~\cite{oddsat}.

A natural difficulty in detecting 
odderon effects in inclusive measurements is the fact that, 
in general, the odderon exchange yields only a small correction to the 
dominating  pomeron contribution to the scattering amplitude. 
On the other hand, this difficulty can be overcome in some cases
by studying the charge asymmetries caused by simultaneous pomeron and 
odderon exchanges \cite{chargeasym}. This measurement looks rather promising
but it was not performed yet, and to this day the best, but still weak, 
experimental evidence for the odderon was found as a difference between 
the differential elastic cross sections for  $pp$ and $p\bar{p}$ scattering 
in the diffractive dip region at $\sqrt{s}=53$ GeV at the 
CERN ISR~\cite{Breakstone}. For a detailed review of the phenomenological 
and theoretical status of the odderon we refer the reader to 
Ref.~\cite{Ewerz-review}.

In the present paper, we study 
the exclusive production of
an  heavy vector meson, 
$V=J/\psi,\, \Upsilon$,  in $pp$ and $p\bar{p}$ collisions: 
$pp\,(\bar{p})\,\rightarrow \, p' \, V \,p''\,(\bar{p}''\,)$;
for a recent review of meson hadroproduction see e.g.~\cite{Lansberg}.
We consider the production of the meson in the central rapidity region, 
separated (in  rapidity) from the two outgoing hadrons $p'$ and 
$p\,''\,(\bar p\,'')$ by two rapidity gaps. 
The vector meson results thus from   pomeron--odderon fusion. 
The mass of the heavy vector meson supplies the hard scale in the process 
of fusion, which may justify a description of the pomeron and the odderon 
within perturbative QCD. The above contribution competes naturally with 
the production of the meson in  pomeron--photon fusion, which is, however, 
under much better theoretical control.

Diffractive production of  the $J/\psi$ meson in proton-(anti)proton 
collisions via pomeron--odderon fusion was investigated already 
in Ref.~\cite{Schafer} in the framework of Regge theory. 
The potential contribution of the $\omega$ reggeon to this process is 
expected to be strongly suppressed due to the Zweig rule.
The estimate of the total $J/\psi$ production cross 
section to be of the order of $75$ nb is quite 
encouraging\footnote{This result does not take the pomeron--photon fusion 
contribution into account.}
.

In this paper we estimate using perturbative QCD the pomeron--odderon and 
pomeron--photon contributions to the exclusive $J/\psi$ and $\Upsilon$ 
hadroproduction, assuming the Tevatron and the LHC conditions.
We find that the exclusive heavy vector meson production in $pp$ and 
$p\bar p$ collisions may serve as a useful tool in  odderon searches. The 
resulting cross sections for  pomeron--odderon fusion are large enough to 
yield large production rates already at the Tevatron for the $J/\psi$ and 
for the $\Upsilon$ at the LHC. The ``background'' photon-driven sub-process
is estimated to have a similar cross section to the pomeron--odderon 
contribution, and in order to clearly isolate the odderon one should 
perform a careful analysis of the transverse momentum distributions of the 
outgoing particles.

The structure of the paper is the following. Section~2 contains a summary 
of the kinematics. In Section~3 we derive the scattering amplitudes 
for the two 
mechanisms of meson hadroproduction. Since the calculational technique
which we use is rather well-known, we present  mostly final results, 
whereas technical details are given in the Appendix.  
Section~4 presents our predictions as well as their discussion.

\begin{figure}[t]
\epsfxsize=6cm
\begin{center}
\epsffile{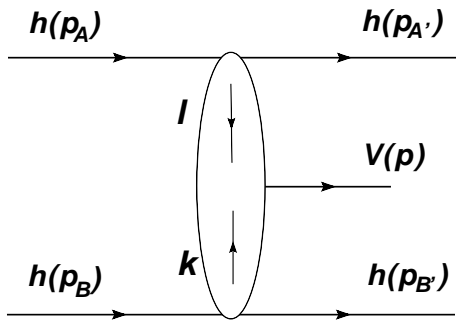}
\end{center}
\caption{\it Kinematics of the exclusive meson production in $pp$ ($p\bar p$) scattering.}
\end{figure}

\section{Kinematics }

We study the processes of hadroproduction shown in Fig.~1,
\begin{equation}
\label{process}
 h(p_A)+h(p_B)\to h(p_{A'}) + V(p) +h(p_{B'}),
\end{equation}
where $h$ and $V$ denote an (anti)proton and a $J/\psi$ (or $\Upsilon$) meson,
respectively. 
In the high-energy limit we neglect the mass of the (anti)proton
$h$ and we identify the momenta $p_A$ and $p_B$ with two light-like Sudakov vectors, 
$p_A^2=p_B^2=0$, so that the scattering energy squared equals 
$s=(p_A+p_B)^2=2p_A\cdot p_B$.

The momenta of the outgoing particles are parametrized as

\begin{equation}
\label{outpart}
p_{A'}=(1-x_{A})p_A + \frac{\bl^2}{s(1-x_A)}p_B -l_\perp \;\;\;\;\mbox{with}\;\;\;\;{\bl}^2=-l_\perp \cdot l_\perp\;,
%\nonumber
\end{equation}
$$
%\begin{equation}
p_{B'}=\frac{\bk^2}{s(1-x_B)}p_A +  (1-x_B)p_B -k_\perp
%\end{equation}
$$
and
$$
%\begin{equation}
p = \alpha_p p_A + \beta_p p_B +p_\perp
%\nonumber
%\end{equation}
$$
\begin{equation}
\alpha_p=x_A - \frac{\bk^2}{s(1-x_B)} \approx x_A \;,\;\;\beta_p=x_B-\frac{\bl^2}{s(1-x_A)}\approx x_B\;,\;\;\;p_\perp = l_\perp + k_\perp\;,
\end{equation}
which lead to the mass-shell condition for the vector meson, 
$V=J/\psi, \Upsilon$, 
\be
\label{mshell}
 m_{V}^2 = s x_A x_B -(\bl +\bk)^2\;.
\ee

\section{The impact-factor representation of scattering amplitudes}

It is well known that at high energies and for small momentum transfers a natural framework to 
calculate the scattering amplitude of the process (\ref{process}) is the $k_\perp$--factorization method, see e.g.\
\cite{Ginzburg}, \cite{eta,eta2,eta3} and references therein. According to this approach, the amplitude is represented as convolutions, over two-dimensional
transverse momenta of $t$--channel partonic reggeons,
 of the impact factors describing scattered nucleons and of the effective production vertex of the vector meson. 
The leading power of $s$ contributing to the scattering amplitude comes from $t$--channel exchanges of gluonic reggeons.

\begin{figure}[t]
\begin{center}
\epsfxsize=5cm
{\large\bf a)} \epsffile{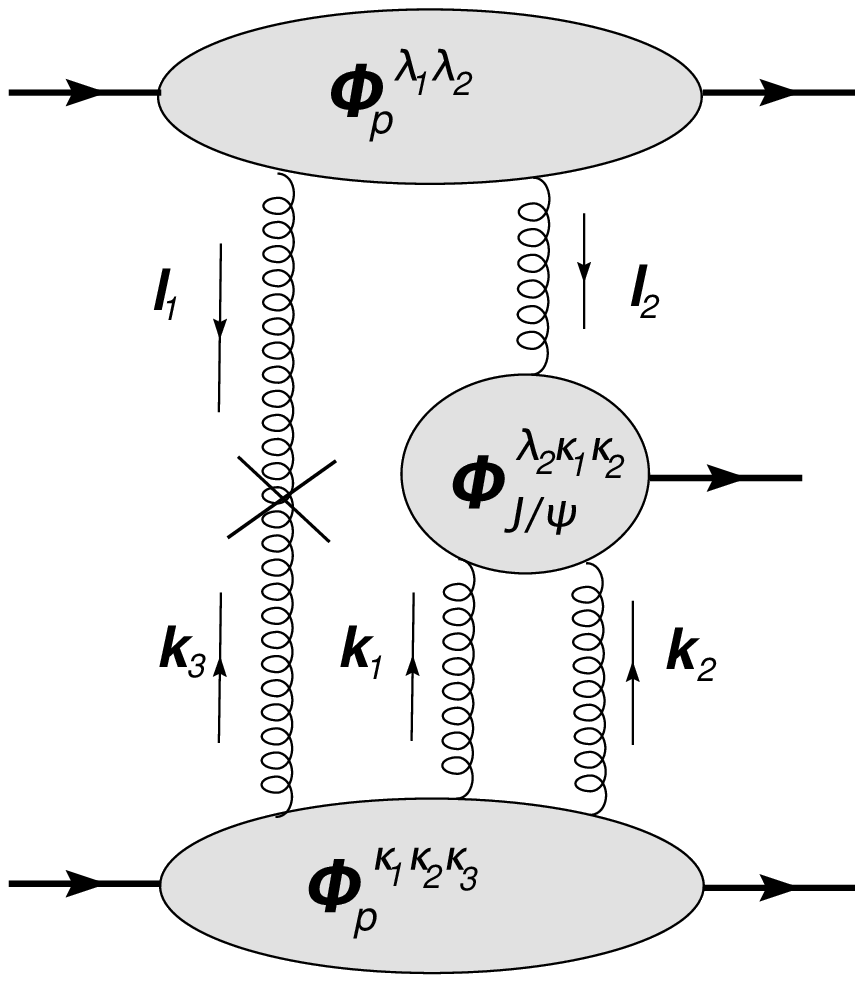} \hspace{1cm}{\large\bf b)}
\epsfxsize=5cm \epsffile{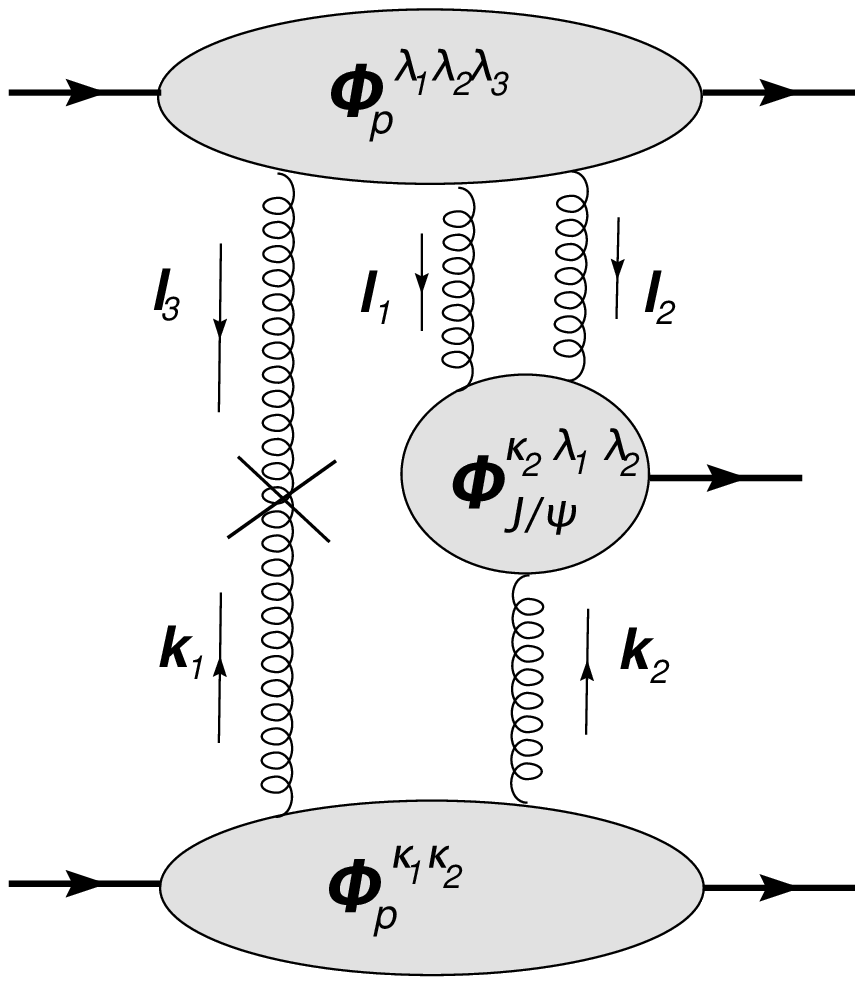}
\end{center}
\caption{\it The lowest order diagrams defining the 
pomeron--odderon fusion amplitudes for
 vector meson production a) ${\cal M}_{P\,O}$ and b) ${\cal M}_{O\,P}$.}
\end{figure}

In the lowest-order approximation, the contributions to the production of 
$J/\psi$ from  pomeron--odderon fusion are shown in Figs.\ 2a,~b. 
The pomeron and the odderon are described in this approximation 
as non-interacting 
longitudinally polarized exchanges  of two and three gluons, respectively. 
The two gluons from the odderon which couple to the effective production vertex 
of the $J/\psi$ will involve the symmetric constants $d^{abc}$ of the colour algebra.
The competing production process of $J/\psi$ from pomeron--photon fusion is
 illustrated in Figs.\ 3a,~b.

Let us first consider  proton--proton scattering. The impact factor representation 
of the diagrams shown in Fig.~2a  reads (see Appendix A1 for details)
\bea
\label{impactA}
&& \cM_{P\;O}=
 \\
&&-is\;\frac{2\cdot3}{2!\,3!}\,\frac{4}{(2\pi)^8}\,\int \frac{d^2 \bl_1}{\bl_1^2}\,\frac{d^2 \bl_2}{\bl_2^2}\ \delta^2(\bl_1 + \bl_2 - \bl)\,\frac{d^2 \bk_1}{\bk_1^2}\,\frac{d^2 \bk_2}{\bk_2^2}\,\frac{d^2 {\bk}_3}{\bk_3^2}\ \delta^2(\bk_1+\bk_2+\bk_3-\bk)
\nonumber \\
&&
\times \delta^2(\bk_3+\bl_1)\,\bk_3^2\;\delta^{\lambda_1 \kappa_3}
\cdot\Phi^{\lambda_1 \lambda_2}_P(\bl_1,\bl_2)\cdot\Phi^{\kappa_1 \kappa_2 \kappa_3}_P(\bk_1,\bk_2,\bk_3)\cdot
\Phi_{J/\psi}^{\lambda_2 \kappa_1 \kappa_2}(\bl_2, \bk_1,  \bk_2)
\nonumber \;.
\eea
Here $\Phi^{\lambda_1 \lambda_2}_P(\bl_1,\bl_2)$ 
denotes the  impact factor of the  proton, scattered via  
pomeron exchange. The gluons forming the pomeron with the momenta $\bl_1$, $\bl_2$ 
carry the colour indices $\lambda_1$, $\lambda_2$, respectively. The corresponding
impact factor of the  proton, scattered via  odderon exchange, is denoted as
 $\Phi^{\kappa_1 \kappa_2 \kappa_3}_P(\bk_1,\bk_2,\bk_3)$. Again,  $\kappa_1$, $ \kappa_2$, $ \kappa_3$ are the colour indices of gluons with the momenta $\bk_1$, $\bk_2$, $\bk_3$. The effective production vertex of the $J/\psi$~meson is
denoted $\Phi_{J/\psi}^{\lambda_2 \kappa_1 \kappa_2}(\bl_2, \bk_1,  \bk_2)$. It results from the fusion of a gluon
with the momentum and the colour index $(\bl_2, \lambda_2)$ from the pomeron with two gluons $(\bk_1,\kappa_1)$ and $(\bk_2,\kappa_2)$ of the odderon. In order to keep the notation of momenta $\bl_i$ and $\bk_j$ 
 most symmetric, we introduced an additional, artificial vertex 
(denoted by the cross in Fig.~2) 
$\delta^2(\bk_3+\bl_1)\,\bk_3^2\;\delta^{\lambda_1 \kappa_3}$
connecting the spectator gluons $(\bl_1,\lambda_1)$ and $(\bk_3,\kappa_3)$. The ratio $\frac{2\cdot3}{2!\,3!}=\frac{1}{2}$ is a combinatorial factor. The factors $\frac{1}{2!}$ and $\frac{1}{3!}$ correct 
the  over-counting of diagrams introduced by factorization in the scattering amplitudes of the
impact factor with pomeron and odderon exchanges, respectively. The factor $2\cdot3=6$ accounts for all possibilities to
build the spectator gluon from the momenta $\bl_i$ and $\bk_j$.

\begin{figure}[t]
\begin{center}
%\\[25mm]
{\large\bf a)}
\epsfxsize=5cm \epsffile{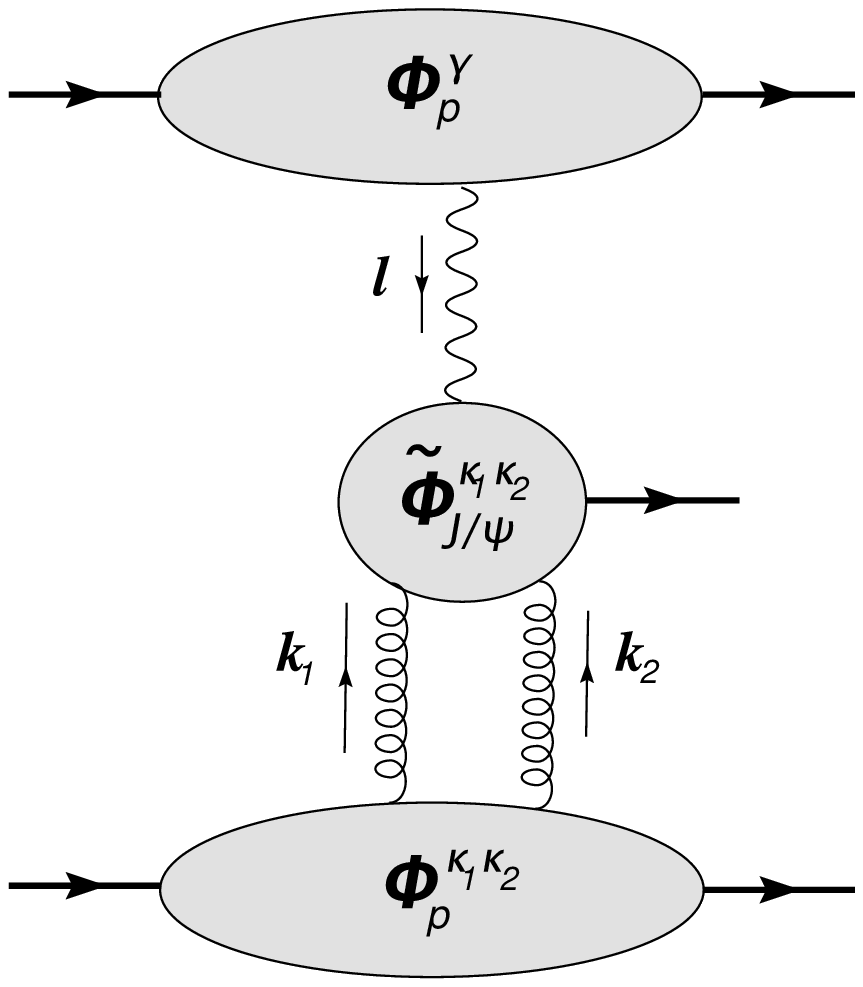} \hspace{1cm}{\large\bf b)}
\epsfxsize=5cm \epsffile{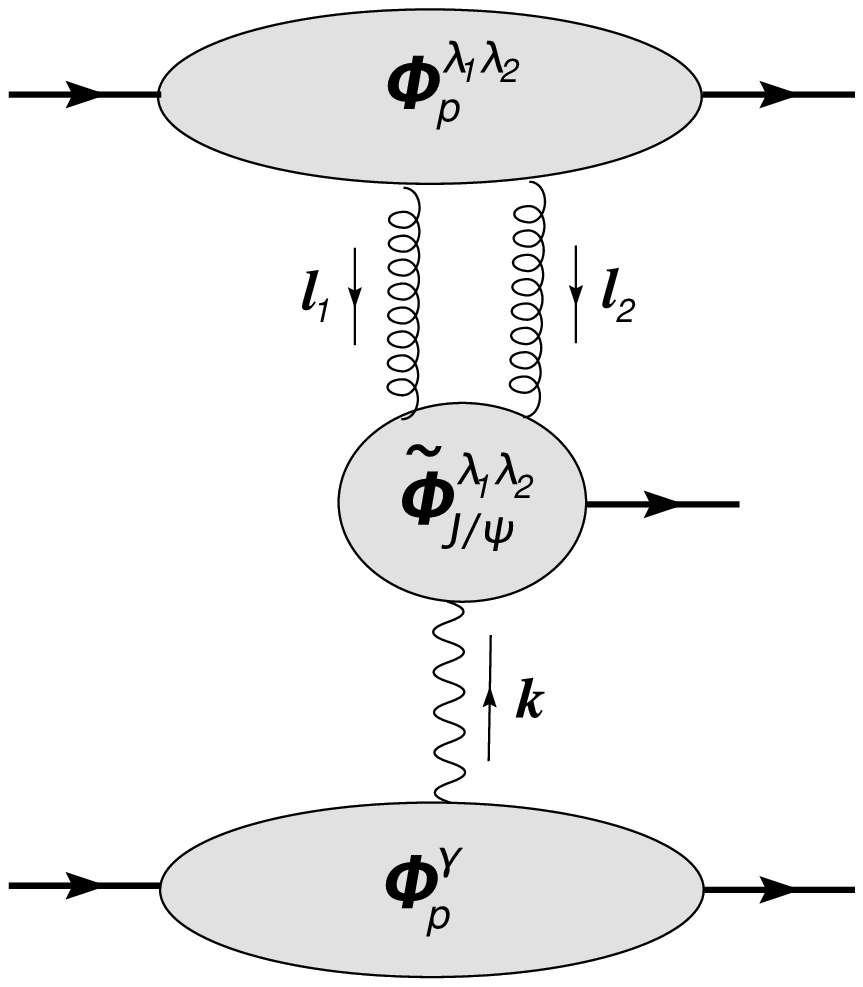} \\
\end{center}
\caption{\it The lowest order diagrams defining the pomeron--photon fusion amplitudes of
the vector meson production a) ${\cal M}_{\gamma\, P}$ and
b)  ${\cal M}_{P\, \gamma}$.}
\end{figure}

The proton impact factors $\Phi^{\lambda_1 \lambda_2}_P(\bl_1,\bl_2)$ and $\Phi^{\kappa_1 \kappa_2 \kappa_3}_P(\bk_1,\bk_2,\bk_3)$ are ``soft", 
non-per\-tur\-ba\-tive objects, therefore to determine their form  we need some non-perturbative 
model of nucleon structure.
In our estimates we use 
 the phenomenological eikonal model of impact factors proposed by
Fukugita and Kwieci\'{n}ski~\cite{Fukugita} (the FK model).  The impact factors can be determined in two steps.  Firstly, the impact factors of a single quark are calculated in the way described in  Refs.~\cite{Ginzburg,eta,eta2,eta3}. Although these calculations are now quite standard, nevertheless in order to make our paper self-contained and to fix the normalization of the impact factors and of the production vertices, we present some technical details in the Appendix. The quark impact factor corresponding to  the pomeron exchange as in Fig.~2a reads (see Appendix A2 for details)
\be
\label{imfaPQ}
\Phi^{\lambda_1 \lambda_2}_q(\bl_1,\bl_2)=-\bar g^2\cdot 2\pi \cdot \frac{\delta^{\lambda_1 \lambda_2}}{2\, N_c}= -\bar \alpha_s\cdot 8\pi^2 \cdot \frac{\delta^{\lambda_1 \lambda_2}}{2\, N_c}\;,
\ee
whereas the corresponding expression with the odderon exchange has the form
\be
\label{imfaOQ}
\Phi^{\kappa_1 \kappa_2 \kappa_3}_q(\bk_1,\bk_2,\bk_3)= i\ \bar g^3\ (2\pi)^2\ \frac{d^{\kappa_3 \kappa_2 \kappa_1}}{4N_c} =i\ \bar \alpha_s^{\frac{3}{2}}\ 2^5\ \pi^\frac{7}{2}\ \frac{d^{\kappa_3 \kappa_2 \kappa_1}}{4N_c} \;,
\ee
with $\bar \alpha_s$ --- the effective coupling constant in the soft region,
 $\bar \alpha_s = \bar g^2/(4\pi)$ and $d^{\kappa_3 \kappa_2 \kappa_1}$ 
 the symmetric structure constants of the colour $SU(3)$ group.
The value of the effective coupling constant  
$\bar \alpha_s$ in Eqs.~(\ref{imfaPQ},\ref{imfaOQ}) is one of the 
main sources of theoretical uncertainties in our estimates and we shall return to this problem in the final discussion.

Secondly, the internal structure of the nucleon is taken into account by``dressing"  the quark impact factors with  
 phenomenological form factors. These form factors should be chosen in a way consistent with the
gauge invariance of QCD, i.e.\ they should vanish when either of momenta $\bl_i$ or $\bk_j$ vanishes.
In the case of  pomeron exchange,  the proton impact factor is modeled as
\be
\label{imfaPom}
\Phi^{\lambda_1 \lambda_2}_P(\bl_1,\bl_2)=3\ {\cal F}_P(\bl_1,\bl_2)\ \Phi^{\lambda_1 \lambda_2}_q(\bl_1,\bl_2)\ \;,
\ee
with
\be
\label{ffPom}
{\cal F}_P(\bl_1,\bl_2)= F(\bl_1+\bl_2,0,0)-F(\bl_1,\bl_2,0)\;,
\ee
where the function $F(\bk_1,\bk_2,\bk_3)$ is taken in the form
\be
\label{F}
F(\bk_1,\bk_2,\bk_3)= \frac{A^2}{A^2 +\frac{1}{2}\left( (\bk_1-\bk_2)^2 + (\bk_2-\bk_3)^2 +(\bk_3-\bk_1)^2 \right)},
\ee
with $A$ being a phenomenological constant chosen to be half of the 
$\rho$~meson mass, $A=m_\rho/2 \approx 384\,$MeV.
The structure of expression (\ref{ffPom}) is quite natural: the first term on the r.h.s.\ of (\ref{ffPom}) corresponds to the
 contribution in which two gluons couple to the same quark line, the second term represents two gluons coupling to two different quarks, 
whereas the factor~3 in (\ref{imfaPom}) counts the number of  valence quarks 
inside the proton.

The corresponding expression for  the proton impact factor with the odderon exchange is constructed in a similar way, as
\be
\label{imfaOdd}
\Phi^{\kappa_1 \kappa_2  \kappa_3}_P(\bk_1,\bk_2,\bk_3)=3\  {\cal F}_O(\bk_1,\bk_2,\bk_3)\ \Phi^{\kappa_1 \kappa_2 \kappa_3}_q(\bk_1,\bk_2,\bk_3)\;,
\ee
where the form-factor ${\cal F}_O$ has the form
\bea
\label{ffOdd}
&&\hspace*{-1cm}{\cal F}_O(\bk_1,\bk_2,\bk_3)=F(\bk=\bk_1+\bk_2+\bk_3,0,0) -\sum\limits_{i=1}^3\ F(\bk_i, \bk-\bk_i,0) +
2\ F(\bk_1,\bk_2,\bk_3)\;,
\nonumber 
\\
&&
\eea
where the function $F$ is defined by Eq.~(\ref{F}).
Again, the first term on the r.h.s.\ of Eq.~(\ref{ffOdd}) corresponds to a contribution when all three gluons couple to a single valence quark, the three terms $F(\bk_i, \bk-\bk_i,0)$ describe the cases when a gluon with 
 momentum $\bk_i$ and two gluons with
 total momentum $\bk-\bk_i$ couple to two different quarks and the last term
describes a coupling of the three gluons to the three different valence quarks of a nucleon.

Let us also note that  
anti-proton impact factors, 
i.e.\  $\Phi^{\kappa_1 \kappa_2 }_{\bar P}$ for  pomeron exchange 
and $\Phi^{\kappa_1 \kappa_2  \kappa_3}_{\bar P}$
for odderon exchange,
are easily obtained from the proton ones: they are given by the same expressions, the only
 modification is the additional 
minus sign for the impact factor of odderon exchange, related to its opposite  charge parity
\be
\label{antiproton}
\Phi^{\kappa_1 \kappa_2 }_{\bar P}=\Phi^{\kappa_1 \kappa_2 }_{ P}\;,\;\;\;\;
\Phi^{\kappa_1 \kappa_2  \kappa_3}_{\bar P}=
-\Phi^{\kappa_1 \kappa_2  \kappa_3}_{ P}\;.
\ee

The derivation of the effective production vertex of a charmonium
$\Phi_{J/\psi}^{\lambda_2 \kappa_1 \kappa_2}(\bl_2, \bk_1,  \bk_2)$ as a part of the impact factor representation
(\ref{impactA})
is one of the main  results of the present study.
For that we assume that the mass $m_{J/\psi}$ of charmonium  supplies a sufficiently hard scale
so we can rely on  perturbation theory. The charmonium
 is treated in the non-relativistic approximation,
where the $\bar c c \to J/\psi$ production vertex has the form
\be
\label{JPsivertex}  
\langle \bar c \ c |J/\psi \rangle = \frac{g_{J/\psi}}{2}\ \hat \varepsilon^{\ *}(p)\left( p \cdot \gamma  + m_{J/\psi}  \right)\;,\;\;\;\;
m_{J/\psi}=2m_c\;,
\ee
where we assume that the $\bar c \ c$ pair is in the colour singlet state, $\varepsilon^*$ is the polarization vector of the 
charmonium.
The coupling constant $g_{J/\psi}$ in (\ref{JPsivertex}) is expressed in terms of the 
electronic width $\Gamma^{J/\psi}_{e^+e^-}$ of the $\,J/\psi \to e^+ e^-$ decay
\be
\label{width}
g_{J/\psi}=\sqrt{ \frac{3m_{J/\psi} \Gamma^{J/\psi}_{e^+e^-}}{16\pi \alpha_{em}^2 Q_c^2} }\;,\;\;\;\;Q_c=\frac{2}{3}\;.
\ee 
The effective production vertex $\Phi_{J/\psi}^{\lambda_2 \kappa_1 \kappa_2}$ as drawn in Fig.~2a can be viewed as
being closely related to the usual impact factor describing the transition of a virtual photon $\gamma^*$ into $J/\psi$ via pomeron exchange. 
Indeed, it is a crossed version of the latter, 
with the $s$--channel $\gamma^*$ replaced by the $t$--channel
gluon of virtuality $-\bl_2^2$ and with two gluons $\bk_2$ and $\bk_3$ in the symmetric $8_S$ colour  representation, instead of  the (also symmetric) 
colour-singlet one. Consequently, the calculation of the $\Phi_{J/\psi}^{\lambda_2 \kappa_1 \kappa_2}$ vertex can proceed in a way analogous to that  of the impact factor of the transition $\gamma^* \to J/\psi$
\cite{Ginzburg}. We thus obtain as a result (technical details of derivation are presented in Appendix A3)
\bea
\label{vertexPO}
&&\Phi_{J/\psi}^{\lambda_2 \kappa_1 \kappa_2}(\bl_2, \bk_1,  \bk_2)=g^3\; \frac{d^{\kappa_1\kappa_2\lambda_2}}{N_c}\;V_{J/\psi}(\bl_2, \bk_1,  \bk_2) =
\alpha_s ^{3\over 2} \, 8\pi^{3\over 2}\; \frac{d^{\kappa_1\kappa_2\lambda_2}}{N_c}\;V_{J/\psi}(\bl_2, \bk_1,  \bk_2), \nonumber
\\
&& \hspace{-1cm}V_{J/\psi}(\bl_2, \bk_1,  \bk_2)=
%\nonumber 
\\
&&4\pi m_c g_{J/\psi}\left[ - \frac{x_B \varepsilon^*\cdot p_B + \varepsilon^*\cdot l_{2\perp}}{\bl_2^2+(\bk_1+\bk_2)^2+4m_c^2} + \frac{\varepsilon^*\cdot l_{2\perp} +\varepsilon^*\cdot p_B\left( x_B - \frac{4\bk_1\cdot \bk_2}{sx_A} \right)}{\bl_2^2+(\bk_1-\bk_2)^2 +4m_c^2}  \right]\;.
\nonumber
\eea
Let us note that, with the mass-shell condition (\ref{mshell}) taken into account, the expression (\ref{vertexPO}) vanishes when either of the momenta $\bl_2$, $\bk_2$ or $\bk_3$ vanishes.  This property is a consequence of the QCD gauge invariance,
that guarantees the infra-red convergence of the 
integrals in the impact-factor 
representation (\ref{impactA}). 

The impact-factor representation of the diagrams shown in Fig.~2b, 
${\cal M}_{O\, P}$, is obtained from the previous formulae by the following 
replacement of the momenta and of the colour indices
\be
\label{OP}
{\cal M}_{O\, P} = {\cal M}_{P \, O}\vert_{\, (\bl_i,\lambda_i) \rightarrow (\bk_i,\kappa_i),\; (\bk_j,\kappa_j)\rightarrow (\bl_j,\lambda_j),\; x_A \leftrightarrow x_B}\;.
\ee

Passing to a description of  $J/\psi$ production in  pomeron--photon 
fusion, the impact-factor representation of the diagrams shown in Fig.~3a reads (see Appendix A1)
\bea
\label{impactC}
&&{\cal M}_{\gamma\, P}=
\\
&&\hspace{-1cm}-\frac{1}{2!} \cdot s \cdot \frac{4}{(2\pi)^4\ \bl^2} \ \Phi^\gamma_P(\bl)\ \int \frac{d^2\bk_1}{\bk_1^2}
\frac{d^2 \bk_2}{\bk_2^2}\ \delta^2(\bk_1+\bk_2-\bk)\ \Phi^{\kappa_1\kappa_2}_P(\bk_1,\bk_2)\ \tilde \Phi^{\kappa_1\kappa_2}_{J/\psi}(\bl,\bk_1,\bk_2)\;.
\nonumber
\eea
Here again, the factor $\frac{1}{2!}$ accounts for the over-counting of diagrams introduced by the factorization of the scattering amplitude involving the
proton impact factor with the pomeron exchange, Eq.~(\ref{imfaPom}).

The photon coupling to the proton involves a phenomenological form factor, which
we take as
\be
\label{gammaff}
\Phi^\gamma_P(\bl)= -i e \cdot F(\bl,0,0)\, .
%\frac{A^2}{A^2+\bl^2}\;.
\ee
It has a proper normalization,  with the $-ie$ coupling, when  $\bl \to 0$. When the 
proton is replaced by an anti-proton, it changes sign
\be
\Phi^\gamma_{\bar P}(\bl) =-\Phi^\gamma_P(\bl)\;,
\ee
similarly to the case of the odderon exchange, Eq.~(\ref{antiproton}).

The effective production vertex of charmonium in  pomeron--photon fusion is, modulo a different colour factor and coupling constants, identical to
the one in  pomeron--odderon fusion (\ref{vertexPO}), see Appendix A3 for details. 
We obtain
\be
\label{vertexGP}
\tilde \Phi^{\kappa_1\kappa_2}_{J/\psi}(\bl,\bk_1,\bk_2)=g^2\ eQ_c\ \frac{2\delta^{\kappa_1\kappa_2}}{N_c}\,
V_{J/\psi}(\bl, \bk_1,  \bk_2)\, = \, \alpha_s \,  eQ_c\, 8\pi\,
\frac{\delta^{\kappa_1\kappa_2}}{N_c}\,V_{J/\psi}(\bl, \bk_1,  \bk_2)\, ,
\ee
with $V_{J/\psi}(\bl, \bk_1,  \bk_2)$ given by Eq.~(\ref{vertexPO}).

Also, let us note that the impact-factor representation of the scattering amplitude corresponding to the diagrams shown in Fig.~3b, 
${\cal M}_{P\, \gamma}$, is obtained from (\ref{impactC}) by the following substitution of  momenta and colour indices
\be
\label{Pg}
{\cal M}_{P\,\gamma} = {\cal M}_{\gamma \, P}\vert_{\, (\bk_i,\kappa_i)\rightarrow (\bl_j,\lambda_j), \; x_A \leftrightarrow x_B}\;,
\ee
analogously to the substitution (\ref{OP}) in the case of pomeron--odderon fusion.

The comparison of the impact-factor representations (\ref{impactA}) and (\ref{impactC}) 
for the two mechanisms of hadroproduction, together with the formulas for the impact factors and the effective vertices,
leads to the conclusion that,  due to different numbers of factors $i$ in both amplitudes, they differ by a relative complex phase factor $e^{i\pi/2}$. It means that the odderon and the photon contributions to the cross section  do not interfere.     

Finally, let us mention that, by replacing $m_{J/\psi}$, $g_{J/\psi}$ and 
$Q_c$ characterizing the charmonium $J/\psi$ by $m_{\Upsilon}$, 
$g_{\Upsilon}$ and $Q_b=1/3$, the formulae of this section describe the 
exclusive hadroproduction of the bottomium $\Upsilon$.

\section{Estimates for the cross section and discussion}

An evaluation of the odderon contribution to the exclusive production cross 
sections of the heavy vector mesons in $pp$ and $p\bar p$ collisions was 
performed numerically. The starting point of this evaluation is the 
amplitude for  pomeron--odderon fusion
\be
{\cal M}_{PO} ^{\mathrm{tot}} = {\cal M}_{PO} +  {\cal M}_{OP}, 
\label{mpo}
\ee
calculated separately for each of the independent polarisation vectors 
$\varepsilon$ of the outgoing vector meson. We focused on an unpolarised 
cross section, so that the cross sections were summed over all 
the polarisations. We consider therefore,
\be
{d \sigma \over dy} = \sum_{\varepsilon} \;
\int_{t^A _{\min}} ^{t_{\max}} \,dt_A\,\int_{t^B_{\min}} ^{t_{\max}} \,dt_B\,
\int_0 ^{2\pi} d\phi  
\, {d \sigma^{(\varepsilon)} \over dy\, dt_A\, dt_B\, d\phi}\; ,
\label{dsdt}
\ee 
where
\be
\label{dsdy}
{d \sigma^{(\varepsilon)} \over dy\, dt_A\, dt_B\, d\phi } = 
{1 \over 512\pi^4 s^2} \, |{\cal M}^{\mathrm{tot}} _{PO}|^2,
\ee
is a differential cross section for the meson polarisation $\varepsilon$,
$t_A=\bl^2$, $t_B=\bk^2$, $\phi$ is the azimuthal angle between $\bk$ and
$\bl$ and $y \simeq \frac{1}{2}\log(x_A/x_B)$ is the rapidity of the
meson in the colliding hadrons c.m.\ frame. 
The lower limits $t^{A} _{\min}$ and $t^{B} _{\min}$ are set to zero for 
 pomeron--odderon fusion. 
The pomeron--photon fusion cross section, $d\sigma_{\gamma} / dy$, 
may be obtained from Eqs.\ (\ref{mpo},\ref{dsdt},\ref{dsdy}) by  
the replacements ${\cal M}_{P\, O} \to {\cal M}_{P\, \gamma}$, 
${\cal M}_{O\, P} \to {\cal M}_{\gamma\, P}$ etc.
The resulting ${d\sigma_{\gamma} \over  dy dt_A dt_B}$, 
however, exhibits the usual 
singular behaviour $\;\sim 1/t_i,\;\; i=A,B$ at $t_i \to 0$, due to the photon 
propagator. A standard kinematic analysis, used e.g.\ in 
the Weizs\"acker--Williams approximation,  provides a lower kinematic 
cut-off on the photon virtuality, giving  
$t^A _{\min} \simeq m_p^2 x_A ^2$ and $t^B _{\min} \simeq m_p^2 x_B ^2$,
with $m_p$ denoting the proton mass (see, e.g.\ \cite{Schuler}).
The upper limit $t_{\max}$ could be, in principle, arbitrarily large, 
but the model of the proton impact factor is  unreliable at larger $t$, 
thus we set $t_{\max} = 1.44$~GeV$^2$.

In the model applied no QCD evolution has been taken into account so far 
and the resulting unpolarised pomeron--odderon differential cross 
section (\ref{dsdy}) does not depend explicitly on the total collision 
energy and on the rapidity of the produced vector meson. 
In order to get reliable predictions for the cross sections this should be
corrected. In what follows, we shall take into account the effects of BFKL 
evolution~\cite{bfkl} and the effects of soft-rescattering which tend to 
destroy the rapidity gap.

We shall include the effects of the BFKL evolution of the 
pomeron using a phenomenological enhancement factor 
$E(s,m_V)$, with $V=J/\psi,\Upsilon$. Note also that the model parameter 
$\bar\alpha_s$ 
enters the pomeron--odderon fusion cross section in the fifth power, 
which may lead to significant uncertainty of the results. 
Thus, for clarity of the discussion, the parameter $\bar\alpha_s$ will be 
explicitly isolated in the presentation of the numerical results.
In addition, the obtained formulae should be corrected for multiple 
soft re-scatterings of proton which can destroy the rapidity gap \cite{dur-chi}. 
Those effects will be expressed as a gap survival factor $S_{\mathrm{gap}}^2$.
Thus, a more realistic cross section, that takes into account necessary 
phenomenological improvements may be written as
\be
\left.
{d \sigma^{\mathrm{corr}} \over dy} 
\right|_{y=0}
\, = \, 
\bar\alpha_s^5\, S_{\mathrm{gap}}^2\, E(s,m_V) \,{d \sigma \over dy},
\label{master}
\ee
where ${d \sigma/ dy}$ is the cross section given by (\ref{dsdt}) at
$\bar\alpha_s=1$.

The calculation is valid only in the high energy limit, 
which implicitly constrains the allowed energy and rapidity range, 
say for $x_A<x_0$ and $x_B<x_0$, and we set $x_0 = 0.1$. 
In numerical evaluations we focus on the central $J/\psi$ and $\Upsilon$ 
production, $y\simeq 0$, where $x_A \simeq x_B \simeq m_V/\sqrt{s}$. 
We approximate the effects of QCD evolution of the pomeron amplitude 
by an exponential enhancement factor $\exp(\lambda \, \Delta y)$ where  
$\Delta y \simeq \log(x_0/x_A)$ is the rapidity evolution length of the 
QCD pomeron. Thus, for the central production one obtains 
\be
E(s,m_V) = (x_0 \sqrt{s}/m_V)^{2\lambda}.
\ee
The effective 
pomeron intercept $\lambda$ depends on the hard scale involved in the 
process (see, e.g. \cite{Navelet}). Following HERA results on the the pomeron 
intercept in exclusive vector meson production we take 
$\lambda = 0.2$ ($\lambda = 0.35$) for the $J/\psi$ ($\Upsilon$) 
production~\cite{h1,zeus}. 
Thus, $E(s,m_V)$ gives a substantial enhancement by a factor of about 5 and 
12 (about 9 and 33) for the $J/\psi$ ($\Upsilon$) production at the 
Tevatron and the LHC correspondingly. 
For the odderon, the rapidity evolution given by the 
Bartels--Kwieci\'{n}ski--Prasza\l{}owicz equation~\cite{BKP} leads 
to a flat dependence on the gap size\footnote{This is true for the 
Bartels--Lipatov--Vacca solution~\cite{BLV} at large rapidities and 
approximately true for the Janik--Wosiek solution~\cite{Wosiek}.}, 
so we neglect the rapidity dependence of the odderon.

Note, that we shall not change the meson production vertex in 
the pomeron--odderon fusion by including into it an (unknown yet) analogue 
of the Sudakov suppression factor for the case of three outgoing gluons.  
An inclusion of the Sudakov-like form-factor would be a desirable improvement 
but the consistent way of taking its effects into account requires 
simultaneously a more detailed analysis of the effects of QCD evolution 
of proton impact factors which is beyond the scope of this paper.

The strong coupling constant in the meson impact factor was set to 
$\alpha_s(m_c) = 0.38$ ($\alpha_s(m_b) = 0.21$), in accordance
with the QCD running. Recall that we assume that $m_c = m_{J/\psi} /2$
and analogously in the case of $\Upsilon$,  $m_b = m_{\Upsilon} /2$.
The available estimates of the effective strong coupling constant, 
$\bar\alpha_s$, of the Fukugita--Kwieci\'{n}ski model, 
yield results with rather large spread. 
The constraints from the data on the total $pp$ and $p\bar p$ 
cross sections gave $\bar\alpha_s = 0.7-0.9$~\cite{Fukugita} 
and a recent thorough analysis of the odderon exchange contribution to the
elastic $pp$ and $p\bar p$ scattering ~\cite{Ewerz-coupling} bounds the 
coupling to be much smaller, $\bar\alpha_s \simeq 0.3$. 
Thus, we performed an independent test of the model based on the vector 
meson photoproduction data. Using the FK model we found the following 
amplitude of $J/\psi$ photoproduction off proton in the forward direction:
\be
\label{mgamp}
{\cal M}_{\gamma} = is \, \pi eQ_c \,\bar\alpha_s \alpha_s(m_c)  \, g_{J/\psi}
\; 
{N_c^2-1 \over N_c^2} \;{ 3\, \log(3 m_c^2 / A^2)\over m_c (m_c^2 - A^2/3)}, 
\ee
and the $t$--dependence (determined numerically) was found to agree reasonably
well with the experimentally measured $\exp(-Bt)$, for moderate $t$, with 
$B\simeq 4.5$~GeV$^{-2}$. Thus, we compared the model estimate of the $J/\psi$ 
exclusive photoproduction cross-section to the data at $W \simeq 10$~GeV,
(equivalent to pomeron $x\simeq x_0$) and we obtained 
$\bar\alpha_s \simeq 0.6-0.7$.

The estimate of uncertainties introduced by $\bar\alpha_s$ and 
$S_{\mathrm{gap}}^2$ should be carried out together. 
The reason for that is that the low value of  $\bar\alpha_s \simeq 0.3$ 
was obtained from an estimate of the odderon exchange in which the soft 
gap survival factor was neglected, thus when it was set 
$S_{\mathrm{gap}}^2=1$. Therefore, for consistency, we shall 
also use  $S_{\mathrm{gap}}^2=1$ in our calculation if the low value of 
$\bar\alpha_s=0.3$ is taken. 
This combination $S_{\mathrm{gap}}^2=1$ and $\bar\alpha_s=0.3$ gives low 
cross-sections and it will be called the {\em pessimistic scenario}.

In the {\em optimistic scenario} we shall use a large value of the
coupling, $\bar\alpha_s = 1$, combined with the gap survival factors 
obtained in the Durham two-channel eikonal model: 
$S_{\mathrm{gap}}^2 =0.05$ for the exclusive production at the Tevatron 
and $S_{\mathrm{gap}}^2=0.03$ for the LHC~\cite{dur-chi,two-channel},
see also \cite{levin}. We believe that the best estimates should follow 
from the {\em central scenario} defined by 
$\bar\alpha_s=0.75$, $S_{\mathrm{gap}}^2 =0.05$ ($S_{\mathrm{gap}}^2=0.03$) 
at the Tevatron (LHC).

\begin{table}[t]
\label{tab1}
\begin{center}%
\begin{tabular}[c]{|c|c|c|c|c|}\hline\hline
$d\sigma/dy$ & 
\multicolumn{2}{c|}{$ J/\psi$} &
\multicolumn{2}{c|}{$\Upsilon$} \\ \cline{2-5}
 &  odderon & photon & odderon & photon    \\ \hline
$p\bar p$ & 20~nb & 1.6~nb &  36~pb & 1.1~pb    \\
$pp$      & 11~nb & 2.3~nb &  21~pb & 1.7~pb    \\\hline \hline
\end{tabular}
\end{center}
\caption{\it Na\"{i}ve cross sections $d\sigma /dy$ given by (\ref{dsdt}) 
for the exclusive $J/\psi$ and $\Upsilon$ production in $pp$ and $p\bar p$ 
collisions by the odderon-pomeron fusion, assuming $\bar\alpha_s=1$ and
analogous cross sections  $d\sigma_{\gamma} /dy$ for the photon 
contribution. The numbers given are partial results only and they
must be improved phenomenologically to provide reliable predictions. }
\end{table}

We analyze the pomeron--photon contribution in a way analogous to
the pomeron--odderon contribution.
In the case of  photon exchange, the $pp$ ($p\bar p$) scatter
typically at large impact parameters and we assume that 
the gap survival $S_{\mathrm{gap}}^2 \simeq 1$ in this case.\footnote{
A more detailed analysis of the gap survival for the photon exchange 
was performed in Ref.~\cite{KMR-phot}. In the same reference a crude estimate
of the pomeron--odderon fusion was obtained, based on the assumption that 
the whole odderon is coupled to the single quark (anti-quark) line.}
Thus, we arrive at the analogue of Eq.~\ref{master} for the photon:
\be
\left. {d \sigma^{\mathrm{corr}}_{\gamma} \over dy}\right|_{y=0} \, = \, 
\bar\alpha_s^2\, E(s,m_V) \,{d \sigma_{\gamma} \over dy}.
\label{masterphot}
\ee

Numerical results for $d \sigma / dy$ and $d \sigma_{\gamma} / dy$ are 
listed in Table\ 1. The photon cross sections depend on the
total collision energy $\sqrt{s}$ through the kinematic dependence
of the lower cut-offs $t^A _{\min}$ and $t^B _{\min}$. 
Thus, the photon cross sections in Table\ 1 for the $p\bar p$ and the $pp$ case 
were obtained assuming the kinematics of the central production at the 
Tevatron ($\sqrt{s}=2$~TeV) and at the LHC  ($\sqrt{s}=14$~TeV) 
respectively. Note that there is a significant difference
between the $pp$ and $p\bar p$ cross sections indicating a significant 
interference between the pomeron--odderon and the odderon--pomeron 
contributions. We stress that the numbers in Table\ 1 represent
only partial results, and they are displayed to provide a basis for
estimates of realistic cross-sections and their uncertainties,
according to the prescription given above.

Besides the cross sections integrated over transverse momenta, we 
calculated also the differential distributions of the produced vector 
mesons, defined as
\be
\left.{d\sigma \over dy d\bp^2}\right|_{\mathrm{norm}} = 
\left( {d\sigma \over dy} \right)^{-1}\times
\sum_{\varepsilon}
\int_{\bk^2<t_{\max}} d^2 k \int_{\bl^2<t_{\max}} d^2 l\, 
{d\sigma^{(\varepsilon)} \over dy\, d^2k \, d^2 l} \, \delta ((\bk+\bl)^2-\bp^2).
\ee
In Fig.~4a and~4b we show the normalised distributions for the $J/\psi$ 
(and the $\Upsilon$) production in $p\bar p$ and $pp$ collisions 
respectively. Clearly, the shapes only weakly depend on the 
vector meson flavour.\footnote{An apparent discrepancy in the normalisation 
of the $J/\psi$ and $\Upsilon$ distributions visible in Fig.~4b emerges 
because we show only part of the $\bp^2$--distributions.}
The production of vector mesons in the forward direction ($\bp^2=0$) 
is maximal for $p\bar{p}$ collisions and vanishes for $pp$ collisions.
This striking difference is caused by an already mentioned interference 
between the pomeron--odderon and the odderon--pomeron contributions.

\begin{figure}[t]
\begin{center}
\noindent
\begin{tabular}{ll}
\parbox{6cm}{
\epsfysize=5cm
\epsfxsize=7cm
\epsffile{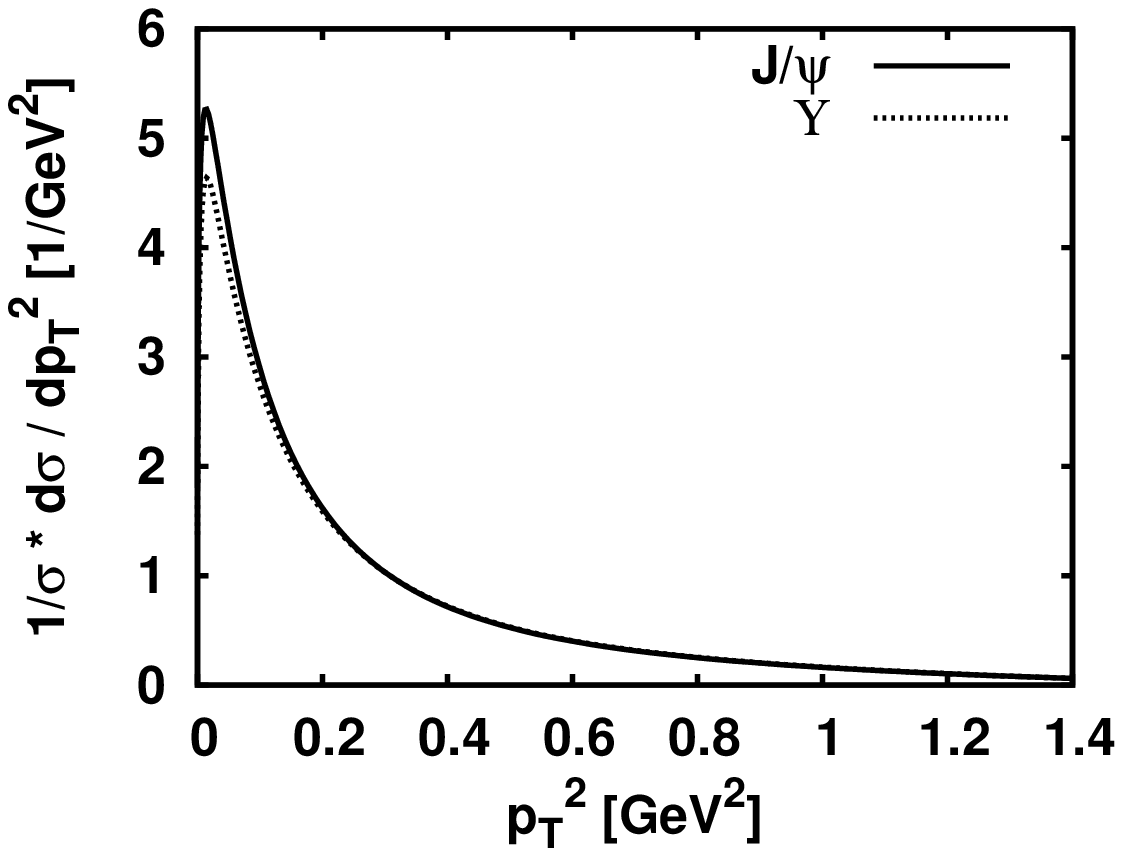} } \hspace{1cm}
&
\parbox{6cm}{
\epsfysize=5cm
\epsfxsize=7cm
\epsffile{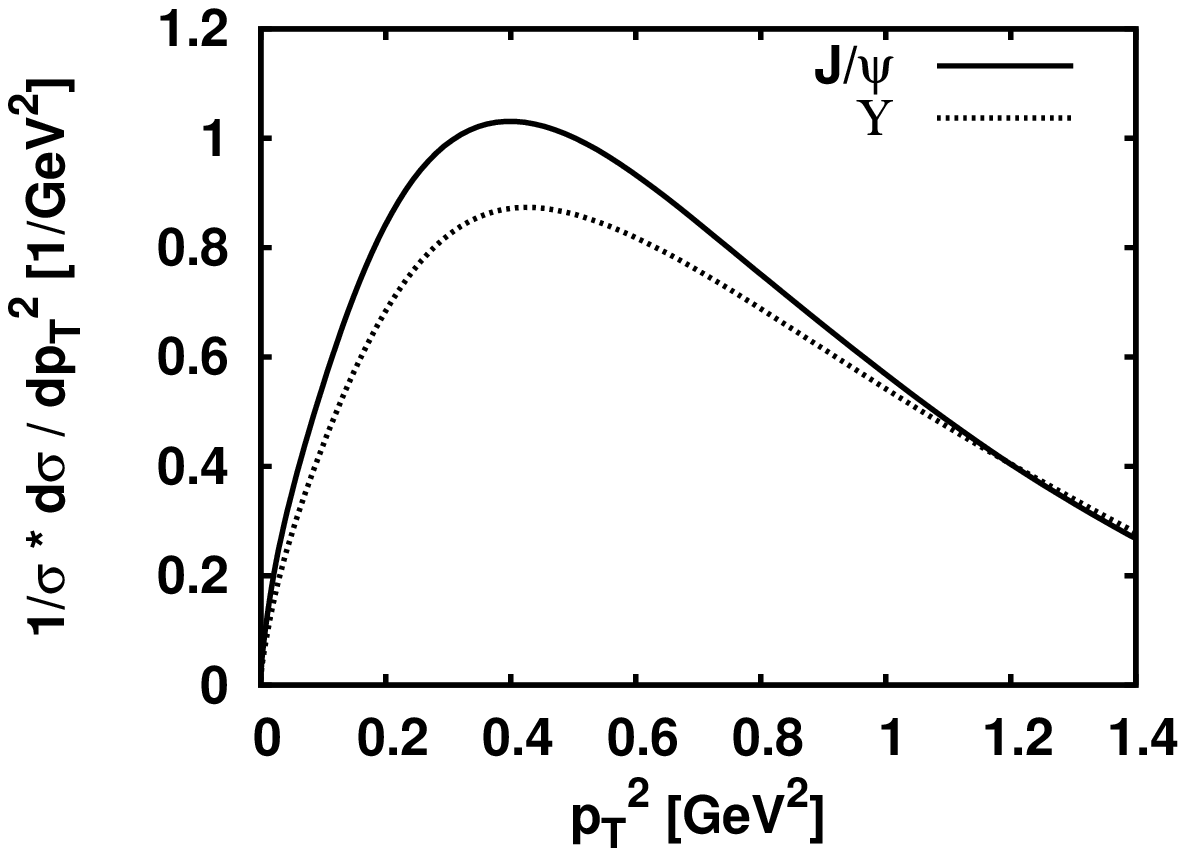} } \\
{\large \bf a)} & {\large \bf b)}
\end{tabular}
\end{center}
\caption{\it $\left. {d\sigma \over dy d\bp^2} \right|_{\mathrm{norm}}\;$ 
for a)  $p\bar p \to p\bar p V$ and b) $pp \to pp V$. }
\end{figure}

The magnitudes of the phenomenologically improved cross sections are
summarized in Table~2. They  were calculated using formulae (\ref{master}) 
and (\ref{masterphot}) accounting for the QCD evolution of the pomeron and the 
gap survival factor, and the uncertainty of $\bar\alpha_s$ was taken into 
account, according to the three scenarios that we consider.   
Recall that the photon and the odderon contributions do not interfere 
in the lowest order approximation and the corresponding cross sections 
may be treated independently.
As seen from the table, the pomeron--odderon contributions are found to be 
uncertain, with a multiplicative uncertainty factor of 3--5. The ambiguities, 
however, cancel partially in the ratio of the pomeron--odderon contribution 
to the pomeron--photon contribution evaluated in the same scenario. 
Thus, within the considered scenarios, the ``odderon to photon ratio'' 
$R=[d\sigma^{\mathrm{corr}} /dy] / [d\sigma^{\mathrm{corr}} _{\gamma}/dy]$
varies between 0.3 and~0.6 for $J/\psi$ production at the Tevatron,
and between about 0.06 and~0.15 at the LHC. 
In the case of $\Upsilon$, $R$~varies between about 0.8 and 1.7 at 
the Tevatron and  between about 0.15 and 0.4 at the LHC. 
These numbers suggest that the odderon contribution may well 
be of a similar magnitude to the photon contribution at the Tevatron 
and somewhat smaller than the photon contribution at the LHC.

Let us note here that the photon-mediated vector meson hadroproduction
may be calculated in a different manner  using the 
Weizs\"{a}cker--Williams approximation.
The dominance of very low virtualities in the photon propagator 
permits to treat one of the protons as a source (with a suitable form factor)
of quasi-real photons that collide with the other proton and produce 
the vector mesons~\cite{KMR-phot,nk1}. In this approximation, the quasi-real 
photon flux is convoluted with a cross-section of the meson photoproduction 
off the proton. The $J/\psi$~photoproduction was measured rather accurately at 
HERA~\cite{h1,zeus} and one may use parametrisations of HERA data to
perform necessary extrapolations. In this approach theoretical uncertainties
and model dependencies are greatly reduced. Thus, calculations based on the 
Weizs\"{a}cker--Williams approximation combined with fits to the HERA data 
give $d\sigma / dy (p\bar p \to p\bar p J/\psi)|_{y=0} 
\simeq 2-2.5$~nb~\cite{KMR-phot,nk1}, somewhat lower than our central 
scenario. For the $\Upsilon$ production at the LHC, 
predictions of Ref.~\cite{nk1}: 
$d\sigma / dy (p p \to p p \Upsilon)|_{y=0} \simeq 100$~pb are
larger than ours
 by a factor of 
more than three. This suggests that  the odderon 
exchange predictions for the $\Upsilon$ production may also be underestimated 
in the central scenario.

%This follows from two reasons. First, within
%the model we use (see Eq.\ \ref{mgamp}), experimental results on the 
%$\Upsilon$ photoproduction are underestimated (that seems to be a typical 
%problem of perturbative estimates of $\Upsilon$ photoproduction). 
%Secondly, we assume a somewhat smaller value 
%of the pomeron intercept $\lambda=0.35$ than the value $\lambda \simeq 0.43$ 
%taken in Ref.~\cite{nk1} for the $\Upsilon$ photoproduction. These differences
%explain well the discrepancy between the predictions. Recall, however, 
%that the pomeron evolution factor and (partially) the cross 
%section normalisation are common for the odderon and the photon mediated 
%production. So, it is fair to state that although estimates of 
%the $\Upsilon$ photo-- and hadroproduction cross sections suffer 
%from significant normalisation uncertainties, the uncertainties may partially 
%cancel in the odderon to photon ratio. 

\begin{table}[t]
\begin{center}%
\label{tab2}
\begin{tabular}[c]{|c|c|c|c|c|}\hline\hline
$d\sigma^{\mathrm{corr}}/dy$ & 
\multicolumn{2}{c|}{$ J/\psi$} &
\multicolumn{2}{c|}{$\Upsilon$} \\ \cline{2-5}
 &  odderon & photon & odderon & photon \\ \hline
Tevatron &
%jpsi
0.3--1.3--5~nb & 
0.8--5--9~nb &  
%upsilon
0.7--4--15~pb & 
0.8--5--9~pb  \\
LHC      & 
%jpsi
0.3--0.9--4~nb & 
2.4--15--27~nb  & 
%upsilon
1.7--5--21~pb & 
5--31--55~pb\\\hline \hline
\end{tabular}
\end{center}
\caption{\it Cross sections $d\sigma^{\mathrm{corr}} /dy|_{y=0}$ 
given by (\ref{master}) for the exclusive $J/\psi$ and $\Upsilon$ production 
in $pp$ and $p\bar p$ collisions by the pomeron--odderon  
fusion, and analogous cross sections  
$d\sigma^{\mathrm{corr}} _{\gamma} /dy|_{y=0}$ for the photon contribution
given by (\ref{masterphot}) for the 
pessimistic--central--optimistic scenarios.}
\end{table}

Our calculations indicate that the odderon-to-photon ratio tends to be of 
the order of unity or smaller, which makes it difficult to get a clear signal
of the odderon from the integrated cross sections. The ratio, however, may be 
enhanced if suitable cuts on outgoing protons transverse momenta are imposed.
Namely, the photon exchange is dominated by very small photon virtualities
(as it follows e.g.\ from the Weizs\"{a}cker--Williams approximation), 
and, for instance for $t_{A}, t_{B} >0.25$ GeV$^{2}$ the pomeron--odderon 
fusion contribution decreases by about one order of magnitude, being still 
visible, and the pomeron--photon fusion contribution decreases by more than 
two orders of magnitude. Then, the odderon contribution could well be a few 
times larger than the photon contribution. Thus, a careful analysis 
of the outgoing proton momenta distribution should permit clear identification 
of the odderon and the photon contributions.

As a final point, let us indicate briefly the possibility to probe the 
odderon via the $\Upsilon$ hadroproduction at the LHC in an asymmetric 
kinematic situation, using the forward detectors, as for instance the 
planned forward proton spectrometer FP420~\cite{FP420}. 
This detector may be capable
of measuring
 the outgoing proton energy and 
transverse momentum with a very good accuracy, for protons that would lose 
about 1\% of their energy. This corresponds to $x_A \simeq 0.01$ 
(see Section~2). 
For  $\Upsilon$ production in the exclusive process it leads to 
$x_B = m_{\Upsilon}^2 / (sx_A) \simeq 5\cdot 10^{-5}$. 
The bottomonium emerging at the rapidity $y_{\Upsilon} \simeq 2.7$ should be 
possible to detect in the $\mu^+\mu^-$ decay channel, and the proton 
$p_B$ would escape  detection. Clearly, due to the small--$x$ evolution 
of the pomeron, the dominant contribution to the production amplitude 
should then come from the pomeron propagating across the large rapidity gap, 
related to $x_B$, and the odderon or photon should span the smaller rapidity 
gap, given by $x_A$. More precisely, for $x_A \gg x_B$, the amplitudes 
${\cal M}_{O\,P}$ and ${\cal M}_{\gamma\, P}$ shown in Fig.\ 2b and Fig.\ 3a 
respectively are enhanced by the QCD evolution by a 
factor of $(x_A/x_B)^{\lambda}\simeq 6$ with respect to the amplitude 
${\cal M}_{P\, O}$ and ${\cal M}_{P\, \gamma}$. 
Therefore, in this kinematics the proton $p_A$ couples predominantly to 
the odderon and to the photon, and one could use the difference 
in $\bl^2$--dependence of the photon and the odderon exchange to cut on 
the proton momentum $p_{A'}$: $\bl^2>l^2_{\min}$, and filter out partially 
the pomeron--photon contribution.
An additional advantage of the measurement in this asymmetric kinematics 
is that at $y_{\Upsilon}\simeq 2.7$ the pomeron evolution down to 
$x_B$ provides an overall enhancement by a factor of a few of the 
exclusive $\Upsilon$ hadroproduction cross section with respect to the 
central production, leading to comfortably large cross sections, well in 
reach of the LHC.

\section*{Acknowledgments}

We acknowledge useful discussions with J.~Bartels, C.~Ewerz,
K.~Golec-Biernat,  O.~Nachtmann, B.~Pire and  S.~Wallon. 
This work is partly supported by the Polish (MEiN) research 
grants 1~P03B~028~28, N202 060 31/3199, and by the
Fonds National de la Recherche Scientifique (FNRS, Belgium).
L.Sz.\ and L.M.\ acknowledge a warm hospitality extended to them at 
Ecole Polytechnique and at LPT-Orsay.

\begin{appendix}
\renewcommand{\theequation}{\thesection.\arabic{equation}}
\section{Appendix}
\setcounter{equation}{0}
\subsection{Derivation of the impact-factor representation (\ref{impactA}) and (\ref{impactC})}
The sum of the Feynman diagrams describing the  fusion of the pomeron (two gluons with  total momentum $l$) with
the odderon (three gluons with  total momentum k)  is written in the Feynman gauge as
\bea
\label{facPOgeneral}
&& {\cal M}_{P\, O}=-i \frac{1}{2!\ 3!}\int\ \frac{d^4l_1\, d^4l_2}{(2\pi)^4}\delta^4(l_1+l_2-l)\,\frac{d^4k_1\ d^4k_2\ d^4k_3}{(2\pi)^8}\delta^4(k_1+k_2+k_3-k)
\nonumber \\
&& {\cal S}^{\lambda_1\lambda_2}_{\mu_1\mu_2}(A\to A')\frac{(-ig^{\mu_1\mu_1' })}{l_1^2+i\epsilon}\frac{(-ig^{\mu_2\mu_2' })}{l_2^2+i\epsilon}{\cal S}^{\lambda_1\lambda_2; \kappa_1\kappa_2\kappa_3}_{\mu_1'\mu_2'; \nu_1'\nu_2'\nu_3'}(J/\psi)
\\
&& \hspace{3cm}\frac{(-ig^{\nu_1\nu_1'})}{k_1^2+i\epsilon}\frac{(-ig^{\nu_2\nu_2'})}{k_2^2+i\epsilon}\frac{(-ig^{\nu_3\nu_3'})}{k_3^2+i\epsilon}
{\cal S}^{\kappa_1\kappa_2\kappa_3}_{\nu_1\nu_2\nu_3}(B\to B')\;.
\nonumber
\eea
Here ${\cal S}^{\lambda_1\lambda_2}_{\mu_1\mu_2}(A\to A')$ is the $S$--matrix element describing the transition of the hadronic state $A$ into $A'$ through the  exchange of two gluons with  momenta $l_i$, $i=1,2$. The $S$--matrix carries  Lorentz and  colour indices $\mu_i$ and $\lambda_i$, respectively. ${\cal S}^{\kappa_1\kappa_2\kappa_3}_{\nu_1\nu_2\nu_3}(B\to B')$ is
the $S$--matrix element describing the  transition of hadronic state $B$ into $B'$ through the exchange of three gluons with  momenta $k_j$, $j=1,2,3$. It carries also  Lorentz and  colour indices $\nu_i$ and $\kappa_i$, respectively. Finally, 
${\cal S}^{\lambda_1\lambda_2; \kappa_1\kappa_2\kappa_3}_{\mu_1'\mu_2'; \nu_1'\nu_2'\nu_3'}(J/\psi)$ 
is the $S$--matrix element
describing the fusion of the two gluons forming the pomeron with the three gluons forming the odderon which produces the $J/\psi$.
The $S$--matrices in Eq.~(\ref{facPOgeneral}) are connected by the gluonic propagators in the Feynman gauge. The factorization of the scattering amplitude ${\cal M}_{P\, O}$ in terms of the $S$--matrices of different subprocesses is possible by
introducing an overcounting of contributing diagrams which gets compensated by the  combinatorial factor $1/(2!\ 3!)$.

The gluonic fusion which results in the production of $J/\psi$ involves only three gluons 
in the lowest order of  perturbation theory.
It means, that  in
${\cal S}^{\lambda_1\lambda_2; \kappa_1\kappa_2\kappa_3}_{\mu_1'\mu_2'; \nu_1'\nu_2'\nu_3'}(J/\psi)$ one of the two gluons
$l_i$ together with one of three gluons $k_j$ form the spectator gluon, disconnected from the $S$--matrix describing  fusion.
Such spectator gluon can be formed in $2\cdot 3=6$ ways and each of these possibilities contributes equally to the scattering amplitude   ${\cal M}_{PO}$. It means that we can consider only one such choice, e.g. with the spectator formed by 
gluons $l_1$ and $k_3$, and  multiply the corresponding result by $6$. The formula for  ${\cal M}_{PO}$ can be thus put in the form
\bea
\label{facPOspect}
&& {\cal M}_{PO}=-i \frac{6}{2!\ 3!}\int\ \frac{d^4l_1\, d^4l_2}{(2\pi)^4}\delta^4(l_1+l_2-l)\,\frac{d^4k_1\ d^4k_2\ d^4k_3}{(2\pi)^8}\delta^4(k_1+k_2+k_3-k)
\nonumber \\
&& i(2\pi)^4\delta^4(l_1+k_3)g_{\mu_1' \nu_3'}k_3^2\delta^{\lambda_1\kappa_3} {\cal S}^{\lambda_1\lambda_2}_{\mu_1\mu_2}(A\to A')\frac{(-ig^{\mu_1\mu_1' })}{l_1^2+i\epsilon}\frac{(-ig^{\mu_2\mu_2' })}{l_2^2+i\epsilon}
{\cal S}^{\lambda_2 \kappa_1\kappa_2}_{\mu_2'; \nu_1'\nu_2'}(J/\psi)
\nonumber \\
&& \hspace{3cm}\frac{(-ig^{\nu_1\nu_1'})}{k_1^2+i\epsilon}\frac{(-ig^{\nu_2\nu_2'})}{k_2^2+i\epsilon}\frac{(-ig^{\nu_3\nu_3'})}{k_3^2+i\epsilon}
{\cal S}^{\kappa_1\kappa_2\kappa_3}_{\nu_1\nu_2\nu_3}(B\to B')\;.
\eea
Here, 
${\cal S}^{\lambda_2 \kappa_1\kappa_2}_{\mu_2'; \nu_1'\nu_2'}(J/\psi)$ is  the $S$--matrix element of the fusion of gluons
with the momenta $l_2$, $k_1$ and $k_2$. We write also the artificial vertex $i(2\pi)^4\delta^4(l_1+k_3)g_{\mu_1' \nu_3'}k_3^2\delta^{\lambda_1\kappa_3}$ to ensure  the most symmetric notation 
of the different parts of 
expression (\ref{facPOspect})
in the momenta $l_i$, $k_j$.

The formula (\ref{facPOspect}) can be further rewritten by applying standard approximations valid in  Regge kinematics,
i.e. characterising processes occuring at high-energies, with small momentum transfers.
The dominant  contribution in $s$ to the scattering amplitude is obtained from the longitudinal polarizations of the $t$-channel gluons.
It results from the following substitution of  numerators in the gluonic propagators
\be
\label{nonsense}
g^{\mu_i \mu_i '} \to \frac{p_B^{\mu_i}\  p_A^{\mu_i '}}{p_A\cdot p_B}\;, \;\;\;\;\;\;g^{\nu_j \nu_j '} \to \frac{p_A^{\nu_j}\  p_B^{\nu_j '}}{p_A\cdot p_B}\,,
\ee 
and leads to the highest power  of large scalar products $p_A\cdot p_B=s/2$.

We paramatrize all momenta using the Sudakov decompositions
\be
\label{Sudakov}
l_i=\alpha_{l i}p_A - \beta_{l i}p_B +l_{\perp i} \,,\;\;\;\;\;k_j=- \alpha_{k j}p_A + \beta_{k j}p_B +k_{\perp j}\;,
\ee
so that $d^4l_i = p_A\cdot p_B \ d \alpha_{l i}d\beta_{l i}d^2l_{\perp  i}$ and 
$d^4k_j = p_A\cdot p_B \ d \alpha_{k j}d\beta_{k  j}d^2k_{\perp j}$.

In the Regge kinematics, the values of the longitudinal Sudakov parameters of the gluons in the $t$--channels are strongly ordered.
As a result, in the S-matrix   ${\cal S}^{\lambda_1\lambda_2}_{\mu_1\mu_2}(A\to A')$, one can neglect the dependence on
 the paramaters $\alpha_{l i}$, as they are much smaller than the $\alpha$ components of other momenta characterizing the transition
$h(p_A)\to h(p_{A'})$. Similarly, in the S-matrix   ${\cal S}^{\kappa_1\kappa_2\kappa_3}_{\nu_1\nu_2\nu_3}(B\to B') $
one can neglect the dependence on $\beta_{k  j}$. On the other hand, the $S$--matrix ${\cal S}^{\lambda_2 \kappa_1\kappa_2}_{\mu_2'; \nu_1'\nu_2'}(J/\psi)$ depends effectively only on
$\alpha_{l 2} \approx \alpha_p \approx x_A$
and $\beta_{k  1}$, $ \beta_{k  2}$, subject to the condition $\beta_{k  1}+\beta_{k  2} \approx \beta_p \approx x_B$.

In the  high energy limit, the asymptotics of the scattering amplitude ${\cal M}_{P\,O}$ is determined by small values of the 
longitudinal Sudakov parameters. Consequently, the denominators of the gluon propagators are given by contributions 
coming  only from the
transverse components of the momenta 
\be
\label{momenta}
l_i^2 \approx l_{\perp i}^2 = -\bl_i^2\;, \;\;\;\;\;k_j^2  \approx k_{\perp j}^2 = - \bk_j^2\;.
\ee

All the above remarks permit to represent  ${\cal M}_{P\, O}$ as a convolution in  transverse momenta of $t$--channel gluons
\bea
\label{POtrans}
\hspace{-1.5cm}&& {\cal M}_{P\, O}=
\\
&& \hspace{-1.5cm}-is\, \frac{6}{2!\ 3!} 
\frac{4}{(2\pi)^8}
\int\ \frac{d^2\bl_1}{\bl_1^2}\frac{d^2\bl_2}{\bl_2^2}\delta^2(\bl_1+\bl_2-\bl)\,
\frac{d^2\bk_1}{\bk_1^2}\frac{ d^2\bk_2}{\bk_2^2} \frac{d^2\bk_3}{\bk_3^2}\delta^2(\bk_1+\bk_2+\bk_3-\bk)
\nonumber \\
&& \hspace{-1.5cm}\delta^2(\bl_1+\bk_3)\bk_3^2\delta^{\lambda_1\kappa_3} \int d\beta_{l_1} {\cal S}^{\lambda_1\lambda_2}_{\mu_1\mu_2}(A\to A')\frac{p_B^{\mu_1} p_B^{\mu_2}}{s} \int d\alpha_{k_3} d\alpha_{k_1}
{\cal S}^{\kappa_1\kappa_2\kappa_3}_{\nu_1\nu_2\nu_3}(B\to B')\frac{p_A^{\nu_1}p_A^{\nu_2}p_A^{\nu_3}}{s}
\nonumber \\
&& \hspace{7cm}\int d\beta_{k_1}
{\cal S}^{\lambda_2 \kappa_1\kappa_2}_{\mu_2' \nu_1'\nu_2'}(J/\psi)
\frac{p_A^{\mu_2'} p_B^{\nu_1'} p_B^{\nu_2'}}{s}\;,
\nonumber
\eea  
which coincides with Eq.~(\ref{impactA}) if one defines the impact-factor for  pomeron exchange as
\be
\label{imfaP}
\Phi^{\lambda_1 \lambda_2}_P(\bl_1,\bl_2)=
\int d\beta_{l_1} {\cal S}^{\lambda_1\lambda_2}_{\mu_1\mu_2}(A\to A')\frac{p_B^{\mu_1} p_B^{\mu_2}}{s}\;,
\ee
the impact-factor for  odderon exchange as
\be
\label{imfaO}
\Phi^{\kappa_1 \kappa_2 \kappa_3}_P(\bk_1,\bk_2,\bk_3)=
\int d\alpha_{k_3} d\alpha_{k_1}
{\cal S}^{\kappa_1\kappa_2\kappa_3}_{\nu_1\nu_2\nu_3}(B\to B')\frac{p_A^{\nu_1}p_A^{\nu_2}p_A^{\nu_3}}{s}\;,
\ee
and the effective production vertex as
\be
\label{effverPO}
\Phi_{J/\psi}^{\lambda_2 \kappa_1 \kappa_2}(\bl_2, \bk_1,  \bk_2)=
\int d\beta_{k_1}
{\cal S}^{\lambda_2 \kappa_1\kappa_2}_{\mu_2' \nu_1'\nu_2'}(J/\psi)
\frac{p_A^{\mu_2'} p_B^{\nu_1'} p_B^{\nu_2'}}{s}\;.
\ee

It is obvious that an analogous   reasoning  can be applied to the sum of diagrams describing  fusion of the photon with the pomeron in Fig.~3a. The analog of Eq.~(\ref{POtrans}) then reads 
\bea
\label{GPtrans}
&& {\cal M}_{\gamma \, P}= -\frac{s}{2!}\ \frac{4}{(2\pi)^4}\ \frac{\Phi^\gamma_P(\bl)}{\bl^2}
\int \frac{d^2\bk_1}{\bk_1^2}\frac{ d^2\bk_2}{\bk_2^2} \delta^2(\bk_1+\bk_2-\bk)
\nonumber \\
&& \int  d\alpha_{k_1}
{\cal S}^{\kappa_1\kappa_2}_{\nu_1\nu_2}(B\to B')\frac{p_A^{\nu_1}p_A^{\nu_2}}{s} 
\int d\beta_{k_1}
{\cal S}^{ \kappa_1\kappa_2}_{ \nu_1'\nu_2'}(J/\psi)
\frac{ p_B^{\nu_1'} p_B^{\nu_2'}}{s}\;,
\eea
where we introduced the photon coupling to the proton $\Phi^\gamma_P(\bl)$ normalized to the proton charge, $\Phi^\gamma_P(0)=-i\, e$.
Eq.~(\ref{GPtrans}) coincides with the impact-factor representation 
 Eq.~(\ref{impactC})
if  the pomeron--photon effective vertex reads
\be
\label{effverGP}
\tilde \Phi^{\kappa_1\kappa_2}_{J/\psi}(\bl,\bk_1,\bk_2)=\int d\beta_{k_1}
{\cal S}^{ \kappa_1\kappa_2}_{\mu'  \nu_1'\nu_2'}(J/\psi)
\frac{ p_A^{\mu'} p_B^{\nu_1'} p_B^{\nu_2'}}{s}
\ee
and if the definition of the impact-factor for   pomeron exchange (\ref{imfaP}) is used for the transition $h(p_B)\to h(p_{B'})$.

\subsection{Derivation of the quark impact-factors (\ref{imfaPQ}) and (\ref{imfaOQ})    }

The quark impact-factor with exchange of the pomeron is defined by Eq.~(\ref{imfaP}) specified for a quark target.
The $S$--matrix corresponding to this transition is described by two diagrams and their colour singlet 
contribution reads
\bea
\label{derivQP}
&& \int d\beta_{l_1} {\cal S}^{\lambda_1\lambda_2}_{\mu_1\mu_2}(A\to A')\frac{p_B^{\mu_1} p_B^{\mu_2}}{s}
= 
\nonumber \\
&& \hspace{-1cm}-i\bar g^2\ \frac{\delta^{\lambda_1\lambda_2}}{2N_c}\int d\beta_{l_1}\ 
\left( \frac{1}{\beta_{l_1}+i\epsilon} + \frac{1}{-\beta_{l_1}-\frac{\bl^2}{s(1-x_A)}+i\epsilon} \right) = -2\pi\ \bar g^2 
\frac{\delta^{\lambda_1\lambda_2}}{2N_c}\;,
\eea
which reproduces Eq.~(\ref{imfaPQ}).

Similarly,
the quark impact factor with exchange of the odderon is defined by Eq.~(\ref{imfaO}) specified for a quark target.
The $S$--matrix corresponding to this transition is described by six diagrams and their colour singlet 
contribution reads
\bea
\label{derivQO}
&&\int d\alpha_{k_3} d\alpha_{k_1}
{\cal S}^{\kappa_1\kappa_2\kappa_3}_{\nu_1\nu_2\nu_3}(B\to B')\frac{p_A^{\nu_1}p_A^{\nu_2}p_A^{\nu_3}}{s}=
\nonumber \\
&& -i \bar g^3\frac{d^{\kappa_1\kappa_2\kappa_3}}{4N_c}     
\int d\alpha_{k_3} d\alpha_{k_1}
\left( 
\frac{1}{(\alpha_{k_1}+i\epsilon)(\alpha_{k_1} +\alpha_{k_2}+i\epsilon)} +\frac{1}{(\alpha_{k_1}+i\epsilon)(\alpha_{k_1} +\alpha_{k_3}+i\epsilon)} \right.
\nonumber \\
&& \left. + \frac{1}{(\alpha_{k_2}+i\epsilon)(\alpha_{k_2} +\alpha_{k_1}+i\epsilon)} + \frac{1}{(\alpha_{k_2}+i\epsilon)(\alpha_{k_2} +\alpha_{k_3}+i\epsilon)} \right.
 \\
&& \left. + \frac{1}{(\alpha_{k_3}+i\epsilon)(\alpha_{k_3} +\alpha_{k_1}+i\epsilon)} + \frac{1}{(\alpha_{k_3}+i\epsilon)(\alpha_{k_3} +\alpha_{k_2}+i\epsilon)}
\right) = i \bar g^3 (2\pi)^2\ \frac{d^{\kappa_1\kappa_2\kappa_3}}{4N_c}\;,
\nonumber
\eea
where in the last step we used the fact that $\alpha_{k_1} + \alpha_{k_2} +\alpha_{k_3} = - \frac{\bk^2}{s(1-x_B)}$\;.
Expression (\ref{derivQO}) reproduces Eq.~(\ref{imfaOQ}).

\subsection{Derivation of the effective vertices (\ref{vertexPO}) and (\ref{vertexGP})}

\begin{figure}[t]
\centerline{
\epsfysize=6cm
\epsfxsize=11cm
\epsffile{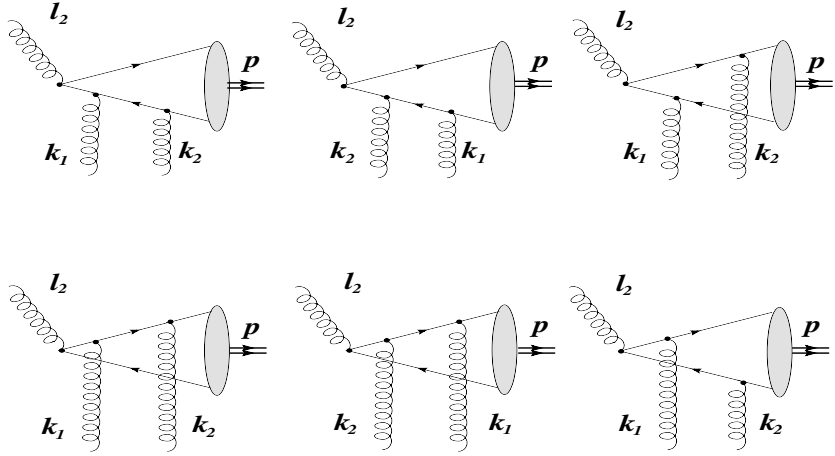} }
\caption{\it The six diagrams defining the effective vertex $g+2g \to J/\psi$.}
\end{figure}

The efective vertex (\ref{effverPO}) is given by the contribution of the six diagrams shown in Fig.~5 with the momenta
$l_2 \approx x_A p_A +l_\perp$, $k_j \approx \beta_{k_j}p_B +k_{\perp j}$, $j=1,2$, where $\beta_{k_1}+\beta_{k_2}\approx x_B$. Taking into account the definition (\ref{JPsivertex}) of the production vertex, the contribution of the 6 diagrams
of Fig.~5 is equal to
\bea
\label{6diag}
&& \hspace{-1.5cm}\int d\beta_{k_1}
{\cal S}^{\lambda_2 \kappa_1\kappa_2}_{\mu_2' \nu_1'\nu_2'}(J/\psi)
\frac{p_A^{\mu_2'} p_B^{\nu_1'} p_B^{\nu_2'}}{s}=-ig^3\  \frac{d^{\lambda_2\kappa_1 \kappa_2}}{2N_c}\  \frac{g_{J/\psi}}{s}
\int d\beta_{k_1}
\nonumber \\
&& \hspace{-1.5cm}\mathrm{Tr}\, \left\{ \left[
\frac{\hat p_A \left( \frac{1}{2}\hat p_{\perp}- \hat l_{2 \perp} +m_c   \right)\hat p_B}{4m_c^2+(\bp-\bl_2)^2 +\bl^2_2}
\left( \frac{1}{\beta_{k_1} -x_B - \frac{2\bk_2^2 -2\bp\cdot \bk_2}{s x_A} +i\epsilon  }    +
\frac{1}{\beta_{k_2} -x_B - \frac{2\bk_1^2 -2\bp\cdot \bk_1}{s x_A} +i\epsilon   }  \right) \right.\right.
\nonumber \\
&& \left. \hspace{-1.5cm}
+ \frac{2}{s^2x_A^2}\hat p_B 
\frac{(\frac{1}{2}\hat p_{\perp}- \hat k_{1 \perp} +m_c )\hat p_A(\hat k_{2 \perp} - \frac{1}{2}\hat p_{\perp} +m_c)
+ (\frac{1}{2}\hat p_{\perp}- \hat k_{2 \perp} +m_c )\hat p_A(\hat k_{1 \perp} - \frac{1}{2}\hat p_{\perp} +m_c)
}{ ( \beta_{k_2} -x_B - \frac{2\bk_1^2 -2\bp\cdot \bk_1}{s x_A} +i\epsilon  ) ( \beta_{k_1} -x_B - \frac{2\bk_2^2 -2\bp\cdot \bk_2}{s x_A} +i\epsilon   )   }
 \hat p_B \right.
\nonumber \\
&& \left. \hspace{-1.5cm}
-\frac{\hat p_B \left( \hat l_{2 \perp} - \frac{1}{2}\hat p_{\perp} +m_c   \right)\hat p_A}{4m_c^2+(\bp-\bl_2)^2 +\bl^2_2}
\left( \frac{1}{\beta_{k_1} -x_B - \frac{2\bk_2^2 -2\bp\cdot \bk_2}{s x_A} +i\epsilon  }    +
\frac{1}{\beta_{k_2} -x_B - \frac{2\bk_1^2 -2\bp\cdot \bk_1}{s x_A} +i\epsilon   }  \right) 
\right] 
\nonumber \\
&& \left. \hspace{8cm}\hat \varepsilon^*\left(  \frac{1}{2}\hat p + m_c   \right) \right\}\;.
\eea
Calculation of the integral over $\beta_{k_1}$, subject to the condition $\beta_{k_1}+\beta_{k_2}\approx x_B$, leads to the result
\be
\label{result6diag}
\int d\beta_{k_1}
{\cal S}^{\lambda_2 \kappa_1\kappa_2}_{\mu_2' \nu_1'\nu_2'}(J/\psi)
\frac{p_A^{\mu_2'} p_B^{\nu_1'} p_B^{\nu_2'}}{s}= g^3\frac{d^{\lambda_2\kappa_1 \kappa_2}}{N_c}V_{J/\psi}(\bl_2,\bk_1,\bk_2)\;,
\ee
which coincides with  Eq.~(\ref{vertexPO}).
\end{appendix}
Finally, let us note that the only difference between the pomeron--photon effective vertex (\ref{vertexGP}) 
and the pomeron--odderon one 
(\ref{vertexPO}) is  the colour factor and the photon coupling. This results in the substitution rule
 $g d^{\lambda_2\kappa_1 \kappa_2} \to 2 eQ_c \delta^{\kappa_1\kappa_2}$, from
 which we recover Eq.~(\ref{vertexGP}).

\end{document}